\documentclass[12pt]{iopart}
\usepackage[utf8]{inputenc}
\usepackage{iopams}
\usepackage{graphicx}

\begin{document}

\title{Schr\"odinger's original quantum-mechanical solution for hydrogen}
\author{Anna Galler$^1$, Jeremy Canfield$^2$ and James K. Freericks$^2$}
\address{$^1$Centre de Physique Th\'eorique, Ecole Polytechnique, Institut Polytechnique de Paris, 91128 Palaiseau Cedex, France\\
$^2$Department of Physics, Georgetown University, 37th and O Sts. NW, Washington, DC 20057, USA}
\ead{james.freericks@georgetown.edu}
\vspace{10pt}
\begin{indented}
\item[]\today
\end{indented}
\begin{abstract}
    In 1926, Erwin Schr\"odinger wrote a series of papers that invented wave mechanics and set the foundation for much of the single-particle quantum mechanics that we teach today. In his first paper, he solved the Schr\"odinger equation using the Laplace method, which is a technique that is quite powerful, but rarely taught. This is unfortunate, because it opens the door to examining quantum mechanics from a  complex-analysis perspective. Gaining this experience with complex analysis is a useful notion to consider when teaching quantum mechanics, as these techniques can be widely used outside of quantum mechanics, unlike the standard Frobenius summation method, which is normally taught, but rarely used elsewhere. The Laplace method strategy is subtle and no one has carefully gone through the arguments that Schr\"odinger did in this first paper, instead it is often just stated that the solution was adopted from Schlesinger's famous differential equation textbook. In this work, we show how the Laplace method can be used to solve for the quantum-mechanical energy eigenfunctions of the hydrogen atom, following Schr\"odinger's original solution, with all the necessary details, and illustrate how it can be taught in advanced instruction; it does require  familiarity with intermediate-level complex analysis, which we also briefly review.
\end{abstract}

\maketitle

\section{Introduction}

In January of 1926, Erwin Schr\"odinger changed the face of physics forever. In his first paper, ``Quantization as an eigenvalue problem (Part I)''~\cite{schroedinger_1926}, Schr\"odinger presents a naive argument to ``derive'' the Schr\"odinger wave equation and then proceeds to solve for the nonrelativistic energy eigenstates of hydrogen, determining the bound-state energy eigenvalues (for $E<0$) and the corresponding unnormalized wavefunctions. He also briefly discusses the continuum solutions with $E>0$. The method he uses is called the Laplace method for solving differential equations. This technique, though similar to the much more familiar Laplace transform, is somewhat different. It is employed to solve differential equations of arbitrary order, but with coefficients that are at most linear functions of the dependent variable. Schr\"odinger relied heavily on the first edition of the differential equation book by Schlesinger~\cite{schlesinger}, which uses this methodology  in its treatment of differential equations. Schr\"odinger is somewhat light on the specific details for how the solution is carried out and, since it is not described in nearly all quantum textbooks, this strategy he used to solve for the first wavefunction has become somewhat of a lost art. In this work, we show the details behind how this first wavefunction was solved in the winter of 1926. As a side note, there has been a lot of historical work on how Schr\"odinger discovered the wave equation. The most comprehensive treatise on this is by Mehra and Rechenberg~\cite{mehra_rechenberg}. So we will not discuss it further here.

Interestingly, the first textbook on ``modern'' quantum mechanics, \textit{The New Quantum Mechanics}, written by George Birtwistle in 1928~\cite{birtwistle} does have an entire section on the Laplace method, but it does not give any additional details beyond what Schr\"odinger gave for solving for the wavefunctions of hydrogen. In general, the Birtwistle book is an uncommon book, reading more like an armchair companion for the original articles than as a true textbook. The 1929 textbook \textit{Quantum Mechanics} by Condon and Morse presents the Frobenius method for solving the hydrogen atom and comments that ``[Schr\"odinger] used the method of complex
integration in arriving at the results which are about to be
obtained here by more elementary methods'' and with that somewhat innocuous comment, so ended the coverage of the Laplace method as a technique for solving the wave equation in nearly all textbooks. Surprisingly, in Landau and Lifshitz's textbook~\cite{landau_lifshitz}, while they use the Frobenius method to solve hydrogen (and many other problems), they use the Laplace method to determine the properties of many of the special functions that appear in quantum mechanics when they cover their mathematical properties in the appendix. These contour methods are briefly covered for the continuum solutions of hydrogen in Bethe and Salpeter's treatise on one and two electron atoms~\cite{bethe_salpeter}, but they also use the Frobenius method for the bound states. There is one other book that we are familiar with that uses the method, Konishi and Paffuti's \textit{Quantum Mechanics}~\cite{konishi_paffuti}, which applies it to the linear potential and the properties of the Airy function in an appendix, but not to the hydrogen problem. The appendix is a rather complete supplement of the mathematical prerequisites needed for this approach.

There also have been some additional discussions of this material in the literature. In 1937, Dirac proposed to generalize momentum from the real axis to the complex plane. He then used this approach to show how to find the wavefunctions of hydrogen, but the treatment has even less details than Schr\"odinger's original paper, although it does appear to correspond to the same solution. More recently, there have been a few papers published, which approach the hydrogen atom from the perspective of the Laplace transform.  Engelfield~\cite{engelfield} describes how to calculate the solutions using the Laplace transform approach (and then the inverse transform involves a contour integral), but it does not give too many details and only examines asymptotic behavior of the wavefunctions. Sherzer~\cite{sherzer} uses a series expansion and term-by-term inverse Laplace transforms to determine the wavefunction. Liu and Mei~\cite{liu_mei} discuss the series method, the Laplace transform method and what they call the transcendental integral method, which is Schr\"odinger's approach via the Laplace method. But again, they provide no details on how the calculation is completed. Tsaur and Wang~\cite{tsaur_wang} use the Laplace transform method to solve a number of different potentials, but they use the definition of the hypergeometric function in terms of the inverse Laplace transform instead of showing details for how the Laplace method is applied to hydrogen.

Part of the motivation for this work is that it is an opportunity to bring complex analysis in a meaningful way into the quantum curriculum for graduate students and help them learn how complex analysis is a powerful tool in physics. The fact that complex analysis is used for many other problems that arise in physics will give these students an edge when they move into research. To some extent, Dirac's long forgotten 1937 paper advocates for just such an approach.

\section{Preliminaries}
The Laplace method is a general technique for solving arbitrary order differential equations that have constant and linear coefficients of each term in the differential equation. While it is related to the well-known Laplace transform, it is distinctly different from it as well. The history is complex~\cite{deakin}. We attribute the method to Laplace, but it was developed by a number of different mathematicians in the latter half of the 19th century. By the turn of the 20th century it was a well-established technique and it made its way into textbooks on differential equations including the textbook written by Schlesinger~\cite{schlesinger} in 1900, which was influential in Schr\"odinger's first publication of the solution of the quantum mechanical hydrogen atom in 1926.
\subsection{The Laplace method}
The Laplace method is a technique to solve ordinary differential equations with constant and linear coefficients, given by the general form
\begin{eqnarray}
    \label{eq:lin_diff}
    \sum_m(a_m+b_mx)y^{(m)}(x)=0.
\end{eqnarray}
The differential equation can be of arbitrary order $m$, but for quantum-mechanics applications, we are most interested in $m=2$. Fixing $m=2$ for concreteness, we obtain the explicit form
\begin{eqnarray}
    (a_2+b_2x)y''(x)+(a_1+b_1x)y'(x)+(a_0+b_0x)y(x)=0.
\end{eqnarray}
The solution to Eq.~(\ref{eq:lin_diff}) is constructed by introducing integrating factors and is represented in the form
\begin{eqnarray}
    \label{eq:lap_trafo}
    y(x)=\int_Ce^{xz}R(z)dz,
\end{eqnarray}
with the added wrinkle that the integral is over a contour $C$ in the complex $z$ plane; the function $R(z)$ is the integrating factor. The challenge in implementing the Laplace method of solution is choosing the correct contour $C$. Note that Eq.~(\ref{eq:lap_trafo}) has the generic form of an inverse Laplace transform, but the standard treatment for solving differential equations by Laplace transforms is simplified from the more general Laplace method.  

In order to determine the complex function $R(z)$ and the contour $C$, we evaluate the derivatives of $y(x)$ by differentiating under the integral sign
\begin{eqnarray}
    y^{m}(x)=\int_Ce^{xz}z^mR(z)dz
\end{eqnarray}
and requiring the contour $C$ to be chosen in such a fashion that differentiating under the integral sign is a valid mathematical procedure. Next, we substitute 
into the differential equation in Eq.~(\ref{eq:lin_diff}) and obtain
\begin{eqnarray}
    \label{eq:diff_coeff}
    \int_Ce^{xz}\sum_m(a_m+b_mx)z^mR(z)dz=0.
\end{eqnarray}
It is convenient to define the polynomials
\begin{eqnarray}
    P(z)=\sum_ma_mz^m \hspace{2em} \mathrm{and}  \hspace{2em} 
    Q(z)=\sum_mb_mz^m.
\end{eqnarray}
Then we rewrite Eq.~(\ref{eq:diff_coeff}) as 
\begin{eqnarray}
    \label{eq:diff_1}
    \int_Ce^{xz}\left[P(z)+Q(z)x\right]R(z)dz=0.
\end{eqnarray}
In order for Eq.~(\ref{eq:diff_1}) to hold, the integrand needs to be the derivative of a complex-valued function and additionally when the function is evaluated at the endpoints of the contour $C$, those two values are required to be the same; often this identical value is zero if the contour extends to infinity. In particular, if the complex-valued function is analytic, and the integral is over a closed contour, then this condition always holds. But care must be taken when the function has branch cuts in the complex plane.  In equations, we have the requirement that
\begin{eqnarray}
    \label{eq:cond_1}
    P(z)R(z)=\frac{d}{dz}\left[Q(z)R(z)\right],
\end{eqnarray}
then Eq.~(\ref{eq:diff_1}) is rewritten as
\begin{eqnarray}
    \label{eq:cond_cont}
    &\int_C\left(\frac{d}{dz}\left[Q(z)R(z)\right]e^{xz}+Q(z)R(z)\frac{d}{dz}e^{xz}\right)dz\nonumber\\
    &=\int_C\frac{d}{dz}\left[Q(z)R(z)e^{xz}\right]dz = 0.
\end{eqnarray}
This requires us to choose the contour $C$ in such a way that the function 
\begin{eqnarray}
    V(z)=Q(z)R(z)e^{xz}
\end{eqnarray}
has equal values at the endpoints of the contour. Note that more than one contour can satisfy these conditions; indeed, for an order-$m$ equation, we know that exactly $m$ linearly independent solutions are represented by $m$ inequivalent contours.
Now we still need to determine the function $R(z)$.  So we simply integrate Eq.~(\ref{eq:cond_1}). The strategy begins by first dividing Eq.~(\ref{eq:cond_1}) by $Q(z)R(z)$ and recognizing that this produces a logarithmic derivative, as follows:
\begin{eqnarray}
    \frac{P(z)}{Q(z)}=\frac{1}{Q(z)R(z)}\frac{d}{dz}\left[Q(z)R(z)\right]=\frac{d}{dz}\ln[Q(z)R(z)].
\end{eqnarray}
Now evaluate the antiderivative on both sides (the constant can be ignored, because the solution of a homogeneous linear differential equation is determined only up to a multiplicative constant)
\begin{eqnarray}
    \ln[Q(z)R(z)]=\int^{z}\frac{P(z')}{Q(z')}dz',
\end{eqnarray}
which, after exponentiation,  yields $R(z)$ in the following form:
\begin{eqnarray}
    R(z)=\frac{1}{Q(z)}\exp\left(\int^{z}\frac{P(z')}{Q(z')}dz'\right).
\end{eqnarray}
Armed with $R(z)$, we immediately find the solution of the differential equation to be
\begin{eqnarray}
    \label{eq:laplace_final}
    y(x)=\int_Ce^{xz}\frac{1}{Q(z)}\exp\left(\int^{z}\frac{P(z')}{Q(z')}dz'\right)dz,
\end{eqnarray}
where the contour $C$ needs to be chosen so that the condition in Eq.~(\ref{eq:cond_cont}) is fulfilled. As stated before, we expect there to be multiple valid choices for the contour $C$.

\subsection{Complex contour integrals, branch cuts}

The Laplace method requires us to calculate contour integrals in the complex plane, such as in Eq.~(\ref{eq:laplace_final}). In general, the integrand is not a single-valued, analytic function in the entire complex plane, but a multi-valued function, which requires us to introduce branch cuts because the integrand involves noninteger powers. For the integrands we work with here, the branch cuts originate from the need to use logarithms to define how $z$  is raised to an  arbitrary power.

\begin{figure}
    \centering
    \includegraphics[width=6.5cm]{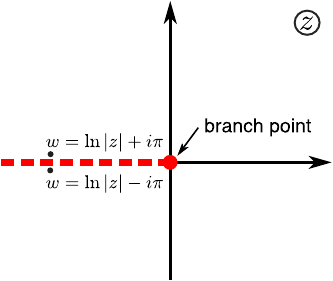}
    \caption{A possible branch cut for $w=\ln{z}$, which runs along the negative real axis.}
    \label{fig:branch_cut}
\end{figure}

The logarithm enters because $z^\alpha$ (with noninteger $\alpha$) is defined to be
\begin{equation}
    z^\alpha=e^{\alpha\ln(z)}.
\end{equation}
The complex logarithm is determined most easily when we represent $z$ in polar coordinates given by the modulus $|z|=\sqrt{zz^*}$ and the polar angle $\phi$, yielding $z=|z|\exp(i\phi)$. We then have
\begin{eqnarray}
    \label{eq:ln_2}
    \ln{z}=\ln(|z|e^{i\phi})=\ln|z|+i\phi.
\end{eqnarray}
From Eq.~(\ref{eq:ln_2}), we can immediately see that $\ln{z}$ is multi-valued---as we move around the origin in a counterclockwise direction, the angle $\phi$ increases from 0 to $2\pi$, so that the imaginary part of the logarithm is not single-valued through the entire complex plane. We need to introduce a branch cut, which can be any curve that does not cross itself and emerges from the origin and goes to infinity; since $\ln(z)=-\ln(1/z)$ implies that infinity is also a branch point of the logarithm. It is common to have the branch cut be a line that goes straight from the origin to infinity and this is what we use here.
The branch cut for $\ln{z}$ drawn  along the negative real axis is shown in Fig.~\ref{fig:branch_cut}. 
The imaginary part of the logarithm is obviously discontinuous across the branch cut. Choosing how we draw the branch cut determines how we define $z^\alpha$.
In particular,
$z^\alpha$ has branch points at $z=0$ and $z=\infty$ that are inherited from the logarithm employed in defining the power. 
Note that no branch points and no branch cut for $z^\alpha$ are needed when $\alpha$ is an integer. In this case, $z^n$ is automatically  single-valued because it is defined in terms of $z$, which is also single valued. 

Taking proper account of how branch cuts are drawn in the complex plane then determines what possible contours we can use for evaluating the contour integrals needed in determining the solutions to the differential equation via Laplace's method. This is detailed below for the solution of hydrogen, but before doing that, we have some more complex analysis we need to cover.

\subsection{$\Gamma$-function and its Hankel representation}
A crucial function needed for asymptotic approximations of contour integrals is Euler's $\Gamma$-function. It is defined from the integral  
\begin{eqnarray}
    \label{eq:g_def}
     \Gamma(p)=\int_{0}^\infty e^{-t}t^{p-1}dt
\end{eqnarray}
for real $p>0$.
For integer arguments we have
\begin{eqnarray}
     \Gamma(n+1)=\int_{0}^\infty e^{-t}t^{n}dt=n! ,
\end{eqnarray}
so that the $\Gamma$-function can be viewed as a function that interpolates the factorial for real positive arguments. Furthermore, by using analytical continuation, one can show that the domain of $\Gamma(p)$ can be extended to the entire complex plane. This is one of the few functions that can be explicitly analytically continued in the complex plane.

Some selected properties of the $\Gamma$-function, which we will need in the following, are the recursive relation
\begin{eqnarray}
    \label{eq:g_rec}
    \Gamma(p+k)=p(p+1)(p+2)...(p+k-1)\Gamma(p)
\end{eqnarray}
and the mirror relation
\begin{eqnarray}
    \label{eq:g_mirror}
    \Gamma(p)\Gamma(1-p)=\frac{\pi}{\sin{(\pi p)}}.
\end{eqnarray}
For a proof of these relations, we refer the reader to Whittaker and Watson~\cite{w_and_w}.

\begin{figure}
    \centering
    \includegraphics[width=8.cm]{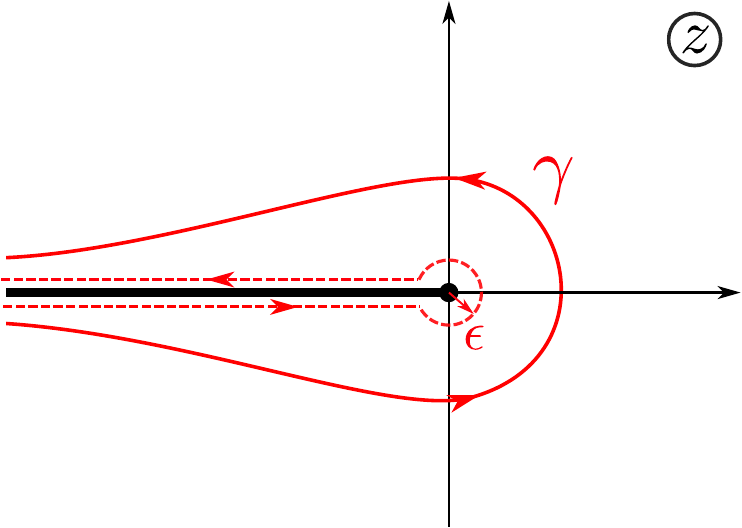}
    \caption{For the calculation of the contour integral in Eq.~(\ref{eq:g_hankel}), we can deform and split the Hankel contour $\gamma$ into three contours: a straight line approaching the branch point $z=0$  from $-\infty$ below the branch cut, a small circle with radius $\epsilon$ around the branch point and another line going towards $-\infty$ above the branch cut.}
    \label{fig:hankel_cont}
\end{figure}

Another important representation of the $\Gamma$-function that we will make use of is its so-called Hankel representation 
\begin{eqnarray}
\label{eq:g_hankel}
    \frac{1}{\Gamma(p)}=\frac{1}{2\pi i}\int_\gamma e^zz^{-p}dz ,
\end{eqnarray}
where $\gamma$ is a Hankel contour shown  in Fig.~\ref{fig:hankel_cont}. To establish that this relation holds, we explicitly evaluate the integral in Eq.~(\ref{eq:g_hankel}). Note that the integral over the Hankel contour would be zero in the case of an analytic single-valued function, however $z^{-p}$ is a multi-valued function (for noninteger $p$) and thus we fix the regular branch so that $z^{-p}$ is real and positive for $z>0$ (on the real axis); we introduce the branch cut from $z=-\infty$ to  $z=0$, as illustrated in Fig.~\ref{fig:branch_cut}. 
To evaluate the integral, we deform the Hankel contour to the dashed contour in Fig.~\ref{fig:hankel_cont} (we can do this because the integrand is analytic in the domain with the branch cut) and we split it into three contour integrals
\begin{eqnarray}
    \label{eq:hankel_split}
    \int_\gamma e^zz^{-p}dz = \lim_{\epsilon\to0}\int_{-\infty}^{-\epsilon} e^zz^{-p}dz+\lim_{\epsilon\to0}\oint e^zz^{-p}dz+\lim_{\epsilon\to0}\int_{-\epsilon}^{-\infty} e^zz^{-p}dz,
\end{eqnarray}
where the first integral is infinitesimally below the negative real axis ($\phi=-\pi$) and the third integral is slightly above ($\phi=\pi$).
We require $\mathrm{Re}~ p<1$ so that the middle integral is finite as $\epsilon\to0$. However, the result is actually valid for all $p$, because the divergence is just an artifact of the dashed-line contour (this apparent divergence can be avoided by shifting the Hankel contour away from the origin). 

With the assumption $\mathrm{Re} ~p<1$, we  evaluate each integral in Eq.~(\ref{eq:hankel_split}) separately.
For the first integral along the real axis below the branch cut, we substitute $z=s e^{-i\pi}$ and calculate
\begin{eqnarray}
    \lim_{\epsilon\to0}\int_{-\infty}^{-\epsilon} e^zz^{-p}dz &= -\int_\infty^0e^{-s}(s e^{-i\pi})^{-p}ds=e^{i\pi p}\int_0^\infty e^{-s}s^{-p}ds\nonumber\\& = e^{i\pi p}\Gamma(1-p).
\end{eqnarray}
In a similar way we find that the integral from $-\epsilon$ to $-\infty$ above the branch cut is
\begin{eqnarray}
    \lim_{\epsilon\to0}\int_{-\epsilon}^{-\infty} e^zz^{-p}dz = -e^{-i\pi p}\Gamma(1-p).   
\end{eqnarray}
The integral along the circle with radius $\epsilon$ around $z=0$ vanishes for $\epsilon\to0$. In fact, with $z=\epsilon e^{i\phi}$, we get
\begin{equation}
    \lim_{\epsilon\to0}\oint e^zz^{-p}dz =  \lim_{\epsilon\to0} i\epsilon^{1-p}\int_{-\pi}^{\pi}e^{i\phi(1-p)}d\phi =0 .
\end{equation}
Combining these results together, we obtain
\begin{eqnarray}
    \frac{1}{2\pi i}\int_\gamma e^zz^{-p}dz &= \frac{1}{2\pi i}\Gamma(1-p)(e^{i\pi p}-e^{-i\pi p}) = \frac{1}{2\pi i}\Gamma(1-p) 2i\sin(\pi p)\nonumber\\ &= \frac{1}{\Gamma(p)},
\end{eqnarray}
where the last step follows from the mirror relation in Eq.~(\ref{eq:g_mirror}).

\section{Laplace method for hydrogen}

\subsection{Schr\"odinger equation for hydrogen}
The nonrelativistic, stationary Schr\"odinger equation for the hydrogen atom in coordinate space reads
\begin{eqnarray}
    \label{eq:schr_orig}
    \mathcal{H}\psi=-\frac{\hbar^2}{2\mu}\nabla^2\psi-\frac{e^2}{ r}\psi=E\psi,
\end{eqnarray}
where $r=\sqrt{x^2+y^2+z^2}$ and $\mu$ is the reduced mass. We can rewrite Eq.~(\ref{eq:schr_orig}) as
\begin{eqnarray}
    \label{eq:schr_full}
    \frac{\partial^2\psi}{\partial x^2}+\frac{\partial^2\psi}{\partial y^2}+\frac{\partial^2\psi}{\partial z^2}+\frac{2\mu}{\hbar^2}\left(E+\frac{e^2}{r}\right)\psi=0.
\end{eqnarray}
 By switching to spherical coordinates, Eq.~(\ref{eq:schr_full}) becomes separable and its solution can be written as
 \begin{eqnarray}
     \psi(r,\theta,\phi)=\chi(r)Y_l^m(\theta,\phi),
 \end{eqnarray}
where $Y_l^m(\theta,\phi)$ is a spherical harmonic and $\chi(r)$ is the solution of the radial equation
\begin{eqnarray}
\label{eq:rad_1}
    \frac{d^2\chi(r)}{dr^2}+\frac{2}{r}\frac{d\chi(r)}{dr}+\left(\frac{2\mu E}{\hbar^2}+\frac{2}{a_0r}-\frac{l(l+1)}{r^2}\right)\chi(r) = 0.
\end{eqnarray}
Here, $a_0=\hbar^2/\mu e^2$ is the Bohr radius, $l$ is the orbital angular momentum quantum number, which can take values $l=0,1,2,3 ...$, and $m$ is the magnetic quantum number, which satisfies $|m|\le l$. The spherical harmonics are simultaneous eigenstates of $\hat{\vec{L}}\,^2$ (with eigenvalue $\hbar^2l(l+1)$) and $\hat{L}_z$ (with eigenvalue $\hbar m$). Note that we are not using Schr\"odinger's original notation here for the integers $n$ and $l$. Schr\"odinger used $n$ to denote the angular momentum quantum number and $l$ to denote the principal quantum number---in modern usage, we do the opposite. In order not to be confusing, we adopt the modern notation throughout this paper.

Equation~(\ref{eq:rad_1}) can have singular behavior when $r\to 0$ and when $r\to\infty$. These points are the boundaries of the domain of the radial coordinate $r$ in the differential equation. Current textbooks state that the wavefunction must be square integrable. In the standard approach, where one writes a differential equation for $r\chi(r)$, the boundary condition assumed is that $r\chi(r)$ vanishes at both boundaries. This condition is equivalent to the requirement that $\chi(r)$ be finite as $r\to 0$ and vanish faster than $1/r$ as $r\to\infty$. In fact, the modern theory of rigged Hilbert spaces actually requires $\chi(r)$ to vanish faster than any power as $r\to\infty$, because it is  bound state. Schr\"odinger stated his requirement that $\chi(r)$ be \textit{endlich}, which means finite. Recall, Schr\"odinger knew nothing about normalization in his first paper, so there was no concept of square integrability in early 1926. Indeed, the requirement that the wavefunction always is \textit{finite} is likely to be the correct condition on wavefunctions, even though textbooks usually use square integrability; as an example consider the two-dimensional particle in a circular box---the Bessel functions with index $m$ are the solutions of the radial equation, yet for $m=0$, the irregular Neumann function is square integrable, but not finite---it must be eliminated since it allows for particle current creation at the origin, hence the condition of a finite wavefunction excludes it, while square integrability does not. As we will see below, the requirement that $\chi(r)$ remains finite in these limits ($r\to 0$ and $r\to\infty$) acts as the boundary condition for the solution and appears to be sufficient for determining the solution.

The differential equation in Eq.~(\ref{eq:rad_1}) is not yet in the form that it can be solved by the Laplace method, because the coefficients are not all linear. We re-express $\chi(r)$ in the form
\begin{eqnarray}
    \chi(r)=r^\alpha U(r)
\end{eqnarray}
Using this expression, the differential equation becomes
\begin{eqnarray}
\label{eq:rad_2}
    U''(r)&+&\frac{2}{r}(\alpha+1)U'(r)+\left(\frac{2\mu E}{\hbar^2}+\frac{2}{a_0r}\right)U(r)\\
    &+&\left(\frac{\alpha(\alpha+1)}{r^2}-\frac{l(l+1)}{r^2}\right)U(r)=0.\nonumber
\end{eqnarray}
The last term vanishes if we choose $\alpha=l$ or $\alpha=-l-1$. 
We choose $\alpha=l$, in order to satisfy the finite condition on $\chi(r)$ with a finite condition on $U(r)$. Requiring that $U$ be finite is a stronger condition than requiring $\chi$ be finite. It is likely that this finiteness condition on $U$ is what Schr\"odinger used in his analysis, but the paper is not completely clear on this point.  In the modern way of handling the differential equation in Eq.~(\ref{eq:rad_1}), we look at an asymptotic analysis as $r\to 0$, which immediately tells us that  $\lim_{r\to 0}U(r)$ must be a nonzero constant. It is also well known, and Schr\"odinger explicitly states, that one can construct the solution when one chooses $\alpha=-l-1$, but it is less convenient, and so we do not show how to do it here.

After this specific choice of $\alpha$, we obtain a differential equation with only linear coefficients, which can now be solved with the Laplace method.
We bring Eq.~(\ref{eq:rad_2}) into a standard form
\begin{eqnarray}
\label{eq:laplace}
    rU''(r)+2(l+1)U'(r)+\left(\frac{2\mu E}{\hbar^2}r+\frac{2}{a_0}\right)U(r)=0
\end{eqnarray}
and construct the corresponding $P(z)$ and $Q(z)$ polynomials (with complex variable $z$)
\begin{eqnarray}
    P(z) &= 2(l+1)z+\frac{2}{a_0} \\
    Q(z) &= z^2+\frac{2\mu E}{\hbar^2} = (z-c_1)(z-c_2),
\end{eqnarray}
which defines the two roots $c_1$ and $c_2$.
After performing the factorization, we find that the roots are
\begin{eqnarray}
    c_1 = +\sqrt{-\frac{2\mu E}{\hbar^2}}, \hspace{4em} c_2 = -\sqrt{-\frac{2\mu E}{\hbar^2}}.
\end{eqnarray}
In this section, we focus on the bound-state solutions, so $E<0$, and this implies that both $c_{1}$ and $c_2$ are real numbers.
The ratio $P(z)/Q(z)$ then becomes
\begin{eqnarray}
    \frac{P(z)}{Q(z)} = \frac{2(l+1)z+\frac{2}{a_0}}{(z-c_1)(z-c_2)} = \frac{\alpha_1}{z-c_1}+\frac{\alpha_2}{z-c_2},
\end{eqnarray}
with
\begin{eqnarray}
    \label{eq:alpha_def}
    \alpha_1 = \frac{\hbar}{a_0\sqrt{-2\mu E}}+l+1
    \end{eqnarray}
and
\begin{eqnarray}
    \alpha_2 = -\frac{\hbar}{a_0\sqrt{-2\mu E}}+l+1.
\end{eqnarray}
We next calculate the antiderivative
\begin{eqnarray}
    \int^z\frac{P(z')}{Q(z')}dz' &= \int^z \left(\frac{\alpha_1}{z'-c_1}+\frac{\alpha_2}{z'-c_2}\right)dz' \nonumber\\&= \ln{[(z-c_1)^{\alpha_1}(z-c_2)^{\alpha_2}]}+c_3
\end{eqnarray}
and choose $c_3=0$ without loss of generality, because it only yields an overall multiplicative constant to the wavefunction. In the next step, we compute the integrating factor
\begin{eqnarray}
    R(z) = \frac{1}{Q(z)}\exp{\left(\int^z\frac{P(z')}{Q(z')}dz'\right)} = (z-c_1)^{\alpha_1-1}(z-c_2)^{\alpha_2-1},
\end{eqnarray}
which allows us to express the solution of the differential equation in Eq.~(\ref{eq:laplace}) in the form of a contour integral in the complex plane:
\begin{eqnarray}
\label{eq:u_gen}
    U(r) = \int_C e^{zr}(z-c_1)^{\alpha_1-1}(z-c_2)^{\alpha_2-1}dz.
\end{eqnarray}
The contour $C$ must be chosen so that
\begin{eqnarray}
\label{eq:c_cond}
    \int_C\frac{d}{dz}\left[e^{zr}(z-c_1)^{\alpha_1}(z-c_2)^{\alpha_2}\right]dz = 0,
\end{eqnarray}
which is equivalent to choosing a closed contour, or having the quantity in the square brackets vanish at the endpoints of the contour.

\subsection{Attempts at a general solution}
For the moment, we exclude the case when $\alpha_1$ and $\alpha_2$ are integer numbers (this case will be discussed in detail in the next section). Then the integrand in Eq.~(\ref{eq:u_gen}) is a multi-valued function and one needs to draw a branch-cut in order to evaluate the integral. Note that the points $c_1$ and $c_2$ are branch points, but unlike the logarithm, which has a branch point at infinity, the integrand here does not, because $\alpha_1+\alpha_2$ is an integer. Hence the branch cut must go from $c_1$ to $c_2$ (but it can do so through infinity).

\begin{figure}
    \centering
    \includegraphics[width=8.cm]{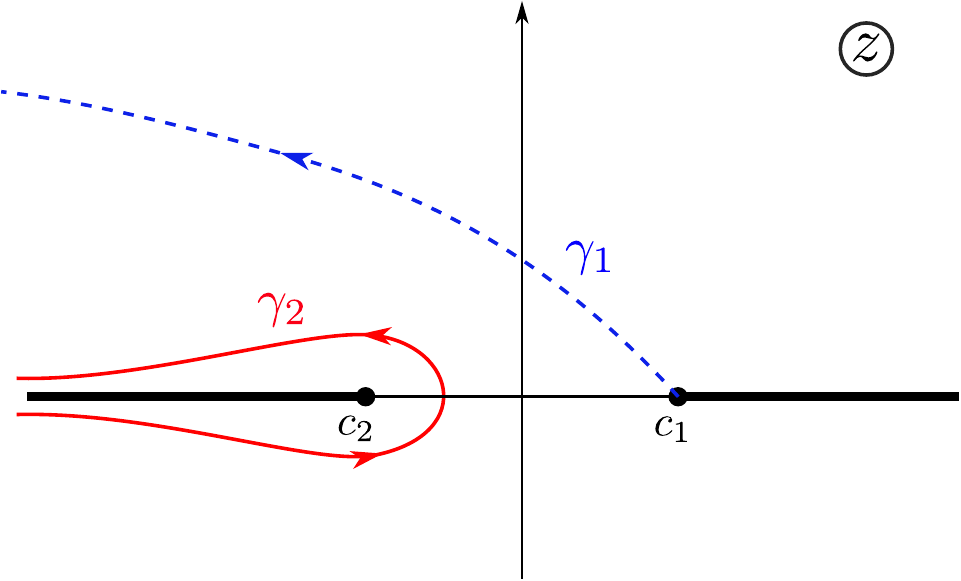}
    \caption{The contours in the complex $z$-plane  $\gamma_1$ and $\gamma_2$ fulfill the condition in Eq.~(\ref{eq:c_cond}) and are employed to calculate $U_1(r)$ and $U_2(r)$.}
    \label{fig:cont_12}
\end{figure}

There is no unique way to draw branch cuts, but they must be chosen consistently in the complex plane. One possible way of introducing a branch-cut is shown in Fig.~\ref{fig:cont_12}. In this case, the branch cut goes from $c_1$ to $c_2$ through the point at $\infty$.
We choose the phase of $z$ to behave like $\exp(i\pi)$ approaching the branch cut that ends at $z=c_2$ from above and $\exp(-i\pi)$ from below. We choose similarly for the branch cut ending at $z=c_1$---the phase is 0 from above and $2i\pi$ from below.

Given those branch cuts, we now consider possible contours $C$ in the complex plane, which fulfill the condition in Eq.~(\ref{eq:c_cond}) about the vanishing of $V(z)$ at the endpoints of the contour. Since $r>0$, we clearly have
\begin{eqnarray}
    \lim_{z\to-\infty}e^{zr} = 0,
\end{eqnarray}
so $z=-\infty$ is always a valid endpoint for a contour (the rigorous condition is that the real part of $z$ approaches $-\infty$).
Other possible endpoints of a contour, where $V(z)$ also vanishes, are when $z=c_1$ and $z=c_2$. Because this is a second-order differential equation, we expect at most two linearly independent solutions, and hence at most two independent contours. 

One possible contour, denoted by $\gamma_1$ in Fig.~\ref{fig:cont_12},  starts at $z=c_1$ and then goes directly towards $-\infty$.
Another possible one, the so-called Hankel contour $\gamma_2$, comes in from $-\infty$ below the branch cut, circles around $z=c_2$ and then goes back to $-\infty$ above the branch cut (see Fig.~\ref{fig:cont_12}). We also could have a contour run from $c_2$ directly to $-\infty$ (above or below the branch cut), but the analysis we give below will rule this contour out as well, so we focus primarily on the Hankel contour here.

To determine whether these contour-integral representations of the solutions of the differential equation are finite, we must do an asymptotic analysis of the integral when $r\to\infty$ and when $r\to 0$. This is equivalent to applying boundary conditions to the solutions of the differential equation.
We now first investigate the contour integral over $\gamma_2$ in the limit $r\to\infty$
\begin{eqnarray}
    U_2(r)=\int_{\gamma_2}e^{zr}(z-c_1)^{\alpha_1-1}(z-c_2)^{\alpha_2-1}dz,
\end{eqnarray}
where we use the subscript 2 to associate this solution with the contour $\gamma_2$.
After a change of variables, given by $z-c_2=s$, we obtain
\begin{eqnarray}
\label{eq:u_s}
    U_2(r)=e^{c_2r}\int_{\gamma_2}e^{sr}(s+c_2-c_1)^{\alpha_1-1}s^{\alpha_2-1}ds,
\end{eqnarray}
where the Hankel contour now encircles the origin in the $s$-plane.
Due to the factor $e^{sr}$, the main contribution to the integral for $r\to\infty$ stems from the vicinity of $s=0$. Therefore, expand $f(s)=(s+c_2-c_1)^{\alpha_1-1}$ into a power series around $s=0$. This yields
\begin{equation}
    f(s)= \sum_{k=0}^\infty\frac{1}{k!}(c_2-c_1)^{\alpha_1-1-k}s^k(\alpha_1-1)(\alpha_1-2) \cdots (\alpha_1-k) 
\end{equation}
or
\begin{eqnarray}
\label{eq:series_2}
    f(s)=\sum_{k=0}^\infty f_ks^k
\end{eqnarray}
with
\begin{eqnarray}
\label{eq:series_coeff}
    f_k=\frac{1}{k!}(c_2-c_1)^{\alpha_1-1-k}(\alpha_1-1)(\alpha_1-2) \cdots (\alpha_1-k).
\end{eqnarray}
By inserting the power series in Eq.~(\ref{eq:series_2}) into the integral in Eq.~(\ref{eq:u_s}), we find 
\begin{eqnarray}
\label{eq:u_s_series}
    U_2(r)\approx e^{c_2r}\sum_{k=0}^N f_k\int_{\gamma_2}e^{sr}s^{\alpha_2+k-1}ds
\end{eqnarray}
under the assumption that we can switch the order of the summation and the integral. It is well known that this analysis yields an asymptotic series, so typically, one only uses a finite number of terms in the final series, because the infinite series will be formally divergent. Nevertheless, a properly truncated asymptotic series can be highly accurate.

Note that $\gamma_2$ is a Hankel contour and each integral term in the series in Eq.~(\ref{eq:u_s_series}) resembles the Hankel-representation of the $\Gamma$-function in Eq.~(\ref{eq:g_hankel}). To express each term in the series in terms of the $\Gamma$-function, we simply perform a change of variables $s=\rho/r$, which then gives us for a general term
\begin{eqnarray}
    \int_{\gamma_2}e^{sr}s^{\alpha_2+k-1}ds &=\int_{\gamma_2}e^{\rho}\left(\frac{\rho}{r}\right)^{\alpha_2+k-1}\frac{d\rho}{r} \nonumber\\ &= r^{-\alpha_2-k}\int_{\gamma_2}e^{\rho}\rho^{\alpha_2+k-1}d\rho \nonumber\\
    &= r^{-\alpha_2-k}\frac{2\pi i}{\Gamma(1-\alpha_2-k)},
\end{eqnarray}
where in the last line we have used the definition of the $\Gamma$-function in Eq.~(\ref{eq:g_hankel}). By making use of the mirror and recursive relations of the $\Gamma$-function, expressed in Eqs.~(\ref{eq:g_mirror}) and (\ref{eq:g_rec}), we can rewrite this term as
\begin{eqnarray}
\label{eq:g_k}
    \int_{\gamma_2}e^{sr}s^{\alpha_2+k-1}ds &= r^{-\alpha_2-k}2i\sin{[(\alpha_2+k)\pi]}\Gamma(\alpha_2+k)\\
    &=r^{-\alpha_2-k}2i(-1)^k\sin{(\alpha_2\pi)}\alpha_2(\alpha_2+1)\cdots(\alpha_2+k-1)\Gamma(\alpha_2).\nonumber
\end{eqnarray}
Inserting this result into Eq.~(\ref{eq:u_s_series}), we obtain
\begin{eqnarray}
\label{eq:asy_ser}
    \lim_{r\to\infty}U_2(r)&\approx e^{c_2r}r^{-\alpha_2}2i\sin{(\alpha_2\pi)}\Gamma(\alpha_2)\nonumber\\
    &\times\sum_{k=0}^\infty(-1)^kr^{-k}f_k\alpha_2(\alpha_2+1)\cdots(\alpha_2+k-1).
\end{eqnarray}
Equation (\ref{eq:asy_ser}) represents an asymptotic series for $U_2(r)$. It does not converge, but for $r\to\infty$ it approximates $U_2(r)$ well with only a few terms.  If we restrict ourselves to the zeroth-order term, we obtain
\begin{eqnarray}
    \label{eq:ur_hankel}
    \lim_{r\to\infty}U_2(r)\approx e^{c_2r}r^{-\alpha_2}2i\sin{(\alpha_2\pi)}\Gamma(\alpha_2)(c_2-c_1)^{\alpha_1-1}.
\end{eqnarray}
Since $c_2<0$, Eq.~(\ref{eq:ur_hankel})  yields a finite result for $U_2(r)$ in the limit $r\to\infty$, consistent with our requirement of a finite wavefunction;  applying a similar analysis with a Hankel contour around $c_1>0$ yields a diverging $U(r)$, which then immediately rules out such a solution. 

We also need to make sure that $U_2(r)$ remains finite for $r=0$.  Setting $r=0$ removes the exponential factor in the integral, yielding
\begin{eqnarray}
    \label{eq:u0_hankel}
    U_2(0)&=\int_{\gamma_2}(z-c_1)^{\alpha_1-1}(z-c_2)^{\alpha_2-1}dz\nonumber\\
    &=\int_{\gamma_2}(s+c_2-c_1)^{\alpha_1-1}s^{\alpha_2-1}ds
\end{eqnarray}
where in the last line we have again used the substitution $z-c_2=s$.
In examining this integral, one can have diverging behavior for $s$ near zero, if $\alpha_2<0$, but this singular behavior can easily be controlled by deforming the Hankel contour to not go too close to $s=0$. 
The integral can also diverge if it does not decay fast enough when $s\to\infty$. On the branch of the Hankel contour that lies below the negative real axis, the phase of $s$ is $-i\pi$ for the term raised to the $\alpha_2-1$ exponent and $i\pi$ for the term raised to the $\alpha_1-1$ exponent according to our choice of branch cuts. But above the branch cut, the phases are both equal to $i\pi$.
In this limit we approximate the integral as the sum of two terms
\begin{eqnarray}
\label{eq:u0_div}
    U_2(0)&\approx&\int_{\gamma_2}s^{\alpha_1-1}s^{\alpha_2-1}ds
    \approx\int_{\infty}^0s^{2l}e^{i\pi(\alpha_1-\alpha_2)}ds+\int^{\infty}_0 s^{2l}e^{i\pi(\alpha_1+\alpha_2)}ds\nonumber\\
    &\approx&\int^{\infty}_0 s^{2l}\left (1-e^{i\pi(\alpha_1-\alpha_2)}\right )ds
\end{eqnarray}
where we used the fact that $\alpha_1+\alpha_2=2l+2$ and the other limit is chosen to be zero for the approximation. The term in parentheses in the last integral is never zero, because we assumed both $\alpha_1$ and $\alpha_2$ were not integers and their difference, which is equal to $2\hbar/(a_0\sqrt{-2\mu E})$, cannot be an even integer. Since the angular quantum number $l$ is a nonnegative integer, we can easily see that the integral in Eq.~(\ref{eq:u0_div}) diverges for $s\to\infty$. Thus the contour integral over $\gamma_2$ yields a diverging $U_2(r)$ for $r=0$, which is clearly not the physical $U(r)$ we are looking for. This divergence is arising entirely from the way we treat the phases in the two different branch cuts, because then the divergence of the integral as $s\to \infty$ cannot be cancelled. Of course a similar analysis rules out any contour originating at $c_2$ and running to $-\infty$ either above or below the branch cut.

Next, we investigate if $\gamma_1$ can yield the correct solution for $U(r)$. This time, we change our integration variable to $s=z-c_1$ and, again, perform a power-series expansion around $s=0$, since in the limit $r\to\infty$ the main contribution to the integral stems from the area close to $s=0$. This yields
\begin{eqnarray}
    U_1(r)&=\int_{\gamma_1}e^{zr}(z-c_1)^{\alpha_1-1}(z-c_2)^{\alpha_2-1}dz\nonumber\\
    &=e^{c_1r}\int_{\gamma_1}e^{sr}s^{\alpha_1-1}(s+c_1-c_2)^{\alpha_2-1}ds,
\end{eqnarray}
where we shifted $s\to s+c_1$ in the second line. Next, we examine the limit as $r\to\infty$ and after using the power series and interchanging the summation with the integral, we find the asymptotic approximation
\begin{equation}
    \lim_{r\to\infty}U(r) \approx e^{c_1r}\sum_{k=0}^N f_k\int_{\gamma_1}e^{sr}s^{\alpha_1+k-1}ds
\end{equation}
by truncating the series after $N$ terms.
The coefficients in this expression can be found exactly and they satisfy
\begin{eqnarray}
    f_k=\frac{(c_1-c_2)^{\alpha_2-1-k}}{k!}(\alpha_2-1)(\alpha_2-2) \cdots (\alpha_2-k).
\end{eqnarray} 
By using a second change of variables, given by $s=e^{i\pi}\rho/r$, we obtain
\begin{eqnarray}
    \lim_{r\to\infty}U_1(r)&\approx e^{c_1r}\sum_{k=0}^N f_k r^{-(\alpha_1+k)}e^{i\pi(\alpha_1+k)}\int_{0}^\infty e^{-\rho}\rho^{\alpha_1+k-1}d\rho \nonumber\\
    &\approx e^{c_1r}r^{-\alpha_1}e^{i\pi\alpha_1}\sum_{k=0}^N f_k(-1)^kr^{-k}\Gamma(\alpha_1+k)
\end{eqnarray}
where in the last line we have used the standard definition of the $\Gamma$-function in Eq.~(\ref{eq:g_def}).
If we restrict ourselves to the zeroth-order term, we obtain
\begin{eqnarray}
\label{eq:u1_div}
    \lim_{r\to\infty}U_1(r)\approx e^{c_1r}r^{-\alpha_1}e^{i\pi\alpha_1}\Gamma(\alpha_1)(c_1-c_2)^{\alpha_2-1}
\end{eqnarray}
which, up to some phase factors, has a similar functional form as $U_2(r)$ in Eq.~(\ref{eq:ur_hankel}).  However, since $c_1>0$, $U_1(r)$ diverges for $r\to\infty$. Hence, this contour $\gamma_1$ cannot be used to define a solution that is finite everywhere.

\begin{figure}
    \centering
    \includegraphics[width=8.cm]{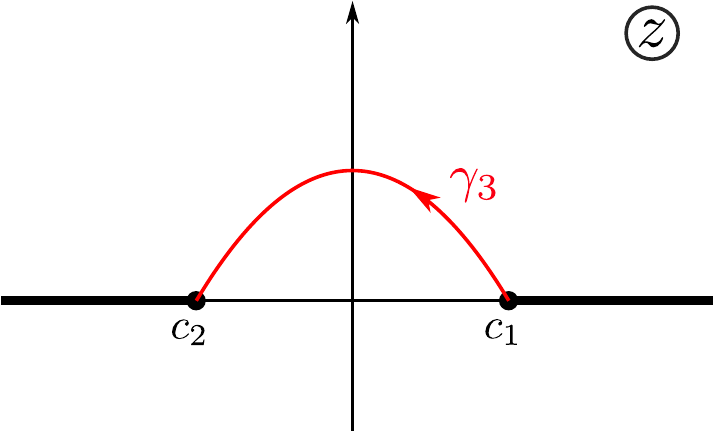}
    \caption{Another possible contour $\gamma_3$ which fulfills the condition in Eq.~(\ref{eq:c_cond}) and can be used to evaluate the integral for $U(r)$.}
    \label{fig:cont_3}
\end{figure}

\begin{figure}
    \centering
    \includegraphics[width=8.cm]{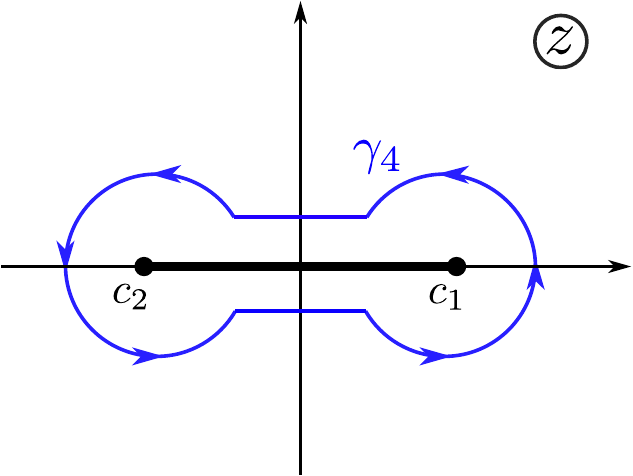}
    \caption{Another possibility of drawing a branch cut and contour $\gamma_4$ in order to evaluate the integral for $U(r)$.}
    \label{fig:cont_4}
\end{figure}

Having not yet found an acceptable solution we consider yet another contour $\gamma_3$, shown in Fig.~\ref{fig:cont_3}, which starts at $z=c_1$ and ends at $z=c_2$. For $r=0$ the contour integral over $\gamma_3$ does not pose any problem. Near $c_1$, the integrand behaves like a power law with a positive exponent. Near $c_2$, the integrand also behaves like a power law and is integrable and finite as long as $\hbar/a_0\sqrt{-2\mu E}<l+2$.  However, in the limit $r\to\infty$, the integral over $\gamma_3$ will pick up a contribution proportional to Eq.~(\ref{eq:u1_div}), with a leading diverging term $e^{c_1r}$. Thus, $U_3(r)$ does not represent the solution we are looking for, either.

Finally, we consider changing the branch cut to one that runs from $c_1$ to $c_2$ directly on the real axis through the origin, as shown in Fig.~\ref{fig:cont_4}. As we wind around the contour $\gamma_4$, the integrand is single valued, because we wind by an angle of $2\pi$ around $c_1$ and around $c_2$, so the integrand winds by an angle of $2\pi (\alpha_1+\alpha_2)$, which is an integer multiple of $2\pi$ and hence is single-valued. 

This new choice of branch cut allows us to draw the new contour $\gamma_4$ circling around both branch points $c_1$ and $c_2$. The asymptotic analysis for such an integral requires us to use the stationary phase approximation. Writing the integrand as an exponential, taking the derivative with respect to $z$, and setting it equal to zero in the limit $r\to\infty$ tell us where the dominant contribution to the integral comes from and allows us to estimate its value. We find that
\begin{equation}
    r+\frac{\alpha_1-1}{z-c_1}+\frac{\alpha_2-1}{z-c_2}=0
\end{equation}
is the stationary phase requirement for all $r$. Taking the limit $r\to\infty$ requires us to analyze the asymptotic behavior near $z=c_1$ and $z=c_2$. This has already been done above, and we find that the point $z=c_1$ leads to an exponentially diverging behavior for $U_4(r)$ as $r\to\infty$.

Thus, none of the possible allowed contour integrals lead to a finite $U(r)$. Indeed, the differential equation in Eq.~(\ref{eq:laplace})  does not have a solution which remains finite for $r\to0$ and $r\to\infty$ for general $\alpha_1$ and $\alpha_2$. The analysis we have given here assumed that $\alpha_1$ and $\alpha_2$ were not integers. The only way to find a finite solution is to examine that case where they are integers. Indeed, this restriction is what leads to energy quantization and the Schr\"odinger wavefunction for the hydrogen atom. We show how to do this next.

\subsection{Solution with quantization}
We now investigate the case when $\alpha_1$ and $\alpha_2$, as defined in Eq.~(\ref{eq:alpha_def}), are real integer numbers. We note that the condition for both of them to be integers is the same, namely  
\begin{eqnarray}
    \label{eq:quant_cond}
    \frac{\hbar}{a_0\sqrt{-2\mu E}} = n \hspace{3em} \mathrm{with} \hspace{1.4em} n=1,2,3,...
\end{eqnarray}
(this quantity is obviously nonnegative and if it equalled zero, then $E\to -\infty$, which we do not allow).
From this we directly obtain for the energy
\begin{eqnarray}
    E_n = -\frac{\hbar^2}{2\mu a_0^2n^2},
\end{eqnarray}
which we recognize as the quantized energy levels in the hydrogen atom, with the principal quantum number $n$. Schr\"odinger also knew these energy levels were correct from experiment, the Bohr model, and Pauli's work, although it is not clear he knew about Pauli's work when he did his work. The condition in Eq.~(\ref{eq:quant_cond}) should thus lead us to the correct solution of the radial wavefunction $U(r)$. First, we note that in the case where $\alpha_{1,2}$ are integers, the integrand in Eq.~(\ref{eq:u_gen}) is not multivalued anymore, but becomes single valued because it is raising the complex monomials $z-c_1$ and $z-c_2$ to integer powers, namely:
\begin{eqnarray}
    \label{eq:u_sv}
    U(r) = \int_C e^{zr}(z-c_1)^{n+l}(z-c_2)^{-n+l}dz.
\end{eqnarray}

This implies that we do not need any branch cuts in the complex plane.  The removal of the branch cuts, allows us to have different options for the contours that satisfy the requirement in Eq.~(\ref{eq:c_cond}).
We now need to distinguish two different cases:  $n\leq l$, where all exponents are nonnegative integers and $n>l$, where one exponent is positive and one is negative.

\begin{figure}
    \centering
    \includegraphics[width=8.cm]{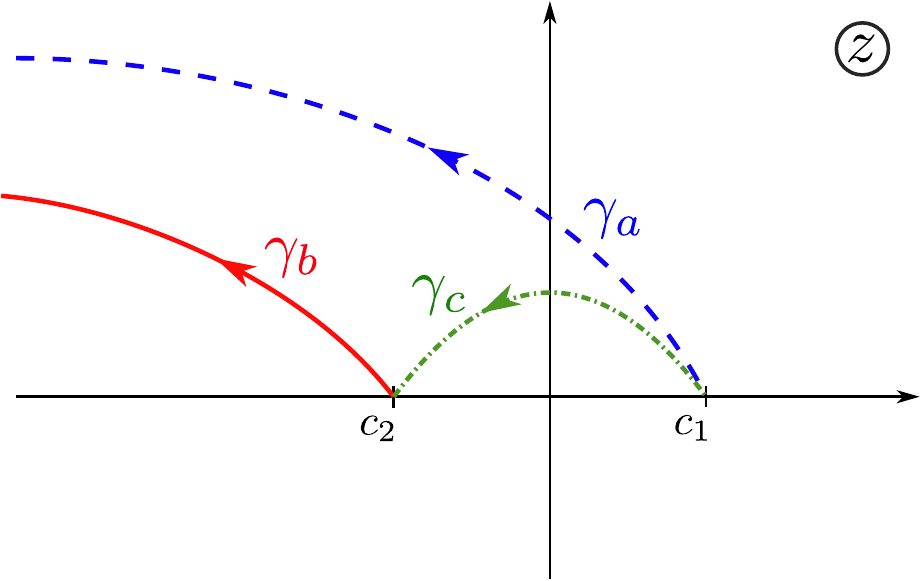}
    \caption{Possible open contours with integer exponents $\alpha_1=n+l$ and $\alpha_2=-n+l$ (see Eq.~(\ref{eq:u_sv})), with $n\leq l$. }
    \label{fig:cont_quant_1}
\end{figure}

We consider first the case with $n\le l$.
This case has no poles in the integrand, anywhere in the complex plane. In fact, the integrand is an analytic function, so any closed contour gives a vanishing result from Cauchy's theorem. The only contours which give a nonzero solution for $U(r)$ are the open contours shown in Fig.~\ref{fig:cont_quant_1}.
However, the divergences we already identified with these contours as $r\to \infty$ or $r\to 0$ remain even in the absence of a branch cut. In particular, for $r\to 0$, $\gamma_a$ and $\gamma_b$  yield a divergence. The integral over $\gamma_c$ is finite for $r=0$,  but diverges as $e^{c_1r}$ in the limit $r\to\infty$ (recall $c_1>0$). We thus do not obtain any physical solution that has a finite $U(r)$ everywhere for $n\leq l$.

\begin{figure}
    \centering
    \includegraphics[width=8.cm]{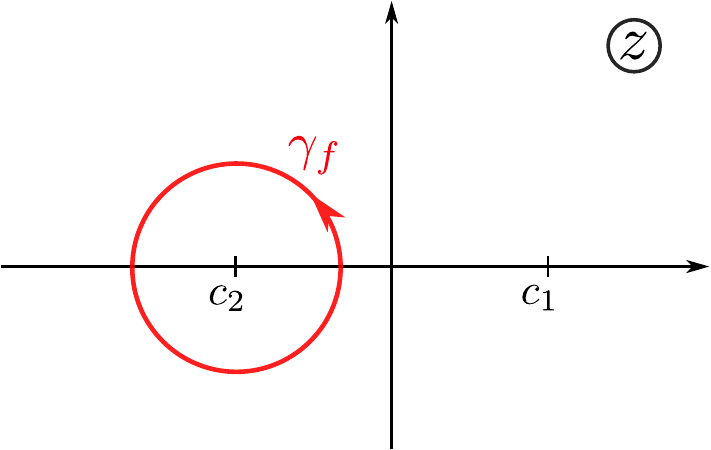}
    \caption{For $n>l$, $z=c_2$ represents a pole of $(n-l)$th order. Thus, one can choose a closed contour enclosing $z=c_2$ to evaluate the integral for $U(r)$ in Eq.~(\ref{eq:u_sv})}
    \label{fig:cont_res}
\end{figure}

We now consider the case $n>l$.
Here, the integrand has a pole at $z=c_2$ of $(n-l)$th order. This allows yet another contour in the complex plane, which yields a non-zero solution for $U(r)$ in Eq.~(\ref{eq:u_sv})---namely a closed contour encircling the pole at $z=c_2$ (see Fig.~\ref{fig:cont_res}). Since this is a closed contour, it automatically satisfies the condition in Eq.~(\ref{eq:c_cond}).
The integral in Eq.~(\ref{eq:u_sv}) can be evaluated by using Cauchy's residue theorem, which yields
\begin{eqnarray}
    \label{eq:u_quant_f}
    U(r) &= \oint_{\gamma_f} e^{zr}(z-c_1)^{n+l}(z-c_2)^{-n+l}dz \nonumber\\
    &= \frac{2\pi i}{(n-l-1)!}\left.\frac{d^{n-l-1}}{dz^{n-l-1}}\left\{e^{zr}(z-c_1)^{n+l}\right\}\right|_{z=c_2}.
\end{eqnarray}
The expression in Eq.~(\ref{eq:u_quant_f}) already resembles the Rodrigues' formula for the Laguerre polynomials, which reads
\begin{eqnarray}
    \label{eq:rodr}
    L_m^{(\alpha)}(y) = \frac{1}{m!}e^yy^{-\alpha}\frac{d^m}{dy^m}\left[e^{-y}y^{m+\alpha}\right].
\end{eqnarray}
In order to bring Eq.~(\ref{eq:u_quant_f}) into this standard form, we use the substitution $(z-c_1)r=-2x$. With this substitution, the point where the derivative in Eq.~(\ref{eq:u_quant_f})  needs to be evaluated, $z=c_2$, becomes $-2x=(c_2-c_1)r=-2c_1r$ and thus $x=c_1r$. With this identification in hand, we obtain
\begin{eqnarray}
    U(r)&= \frac{2\pi i}{(n-l-1)!}(-1)^{n-l-1}r^{n-l-1}
    \left.\frac{d^{n-l-1}}{d(2x)^{n-l-1}}\left\{e^{c_1r}e^{-2x}\left(\frac{-2x}{r}\right)^{n+l}\right\}\right|_{x=c_1r} \nonumber\\
    &= \frac{2\pi i}{(n-l-1)!}(-1)^{-2l-1}r^{-2l-1}e^{c_1r}
    \left.\frac{d^{n-l-1}}{d(2x)^{n-l-1}}\left\{e^{-2x}(2x)^{n+l}\right\}\right|_{x=c_1r} \nonumber\\
    &= 2\pi i(-1)^{-2l-1}(2c_1)^{2l+1}e^{-x}\frac{(2x)^{-2l-1}e^{2x}}{(n-l-1)!}
    \left.\frac{d^{n-l-1}}{d(2x)^{n-l-1}}\left\{e^{-2x}(2x)^{n+l}\right\}\right|_{x=c_1r} \nonumber\\
    &= 2\pi i(-1)^{-2l-1}(2c_1)^{2l+1}e^{-x}L^{(2l+1)}_{n-l-1}(2x)
\end{eqnarray}
by identifying $y=2x$, $m=n-l-1$ and $\alpha=2l+1$.
Thus, by recalling that $\chi(r)=r^lU(r)$, we finally determine the radial solution of the Schr\"odinger equation for the hydrogen atom (up to a constant prefactor):
\begin{eqnarray}
    \chi(x) = x^le^{-x}L^{(2l+1)}_{n-l-1}(2x) \hspace{4em} \mathrm{with} \hspace{3em} x=c_1r=\frac{\sqrt{-2mE}}{\hbar}r.
\end{eqnarray}
We use the definition of the Laguerre-polynomials as given by the sum 
\begin{eqnarray}
    L_n^{\alpha}(x)=\sum_{k=0}^n\frac{(-x)^k}{k!}{{n+\alpha}\choose{n-k}}.
\end{eqnarray}
Using this form, we obtain our final result for the hydrogen wavefunction
\begin{eqnarray}
    \chi(x) = x^le^{-x}\sum_{k=0}^{n-l-1} \frac{(-2x)^k}{k!}{{n+l}\choose{n-l-1-k}},
\end{eqnarray}
which is exactly the form  Schr\"odinger wrote in the original paper in 1926 (except for our interchanging of the integers $n$ and $l$ according to modern nomenclature). When Schr\"odinger completed this work, the concept of normalization of the wavefunction and of the meaning of the wavefunction as a probability amplitude were not yet known. So he did not normalize his final result (although he did discuss normalization as being one way to determine the overall scale of the wavefunction).

\subsection{Continuum solution for hydrogen}

Schr\"odinger did not spend much time discussing the continuum solution and provided no final formulas. Instead, he relied on the Schlesinger solutions and simply stated that there are no issues involved with using the formulas from that book to determine the continuum solutions.  Since the steps needed to carry out this calculation are quite similar to what we have already done with the asymptotic analysis of the contour integrals for the bound states, we will be able to go through the continuum analysis much more quickly and we will complete the study by presenting the continuum wavefunction up to an overall constant. 

\begin{figure}
    \centering
    \includegraphics[width=6.cm]{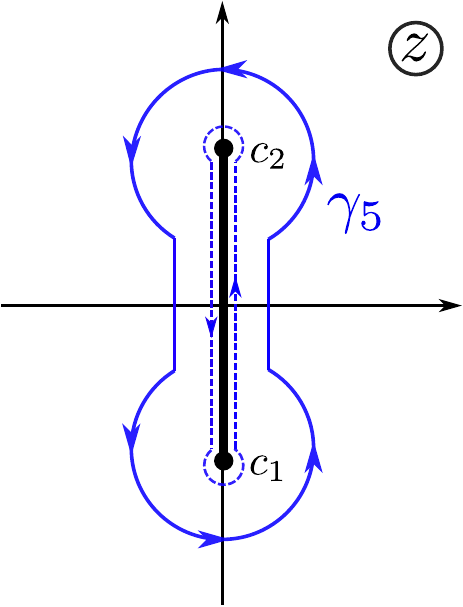}
    \caption{Contour $\gamma_5$ in the complex plane to obtain the continuum solutions with $E>0$. We denote $c_1=-ic$ and $c_2=ic$ with $c=|c_1|$.}
    \label{fig:cont_5}
\end{figure}

The first thing to note is that the constants $c_1$ and $c_2$ are pure imaginary when $E>0$ (for concreteness, we pick $c_1=-ic$ and $c_2=ic$, with $c=|c_1|$). Furthermore, the exponents $\alpha_1$ and $\alpha_2$ are always complex (and are complex conjugates of each other). When determining the possible contours that produce a finite-valued wavefunction, we find as $r\to\infty$, the wavefunction is always finite, because it will behave like $e^{c_1r}$ or $e^{c_2r}$, which are both bounded since the $c_{1,2}$ coefficients are purely imaginary (a more complete analysis is given below). Because the exponents in the integrand become pure imaginary for $l=0$, we cannot have any integral that has $c_1$ or $c_2$ as an endpoint anymore, as the condition in Eq.~(\ref{eq:c_cond}) no longer holds at those points. This means that the only possible contours that work for all $l$ are closed contours or Hankel-like contours, depending on how the branch cuts are chosen. 
Then the only condition that remains is to guarantee that the wavefunction is finite for $r\to 0$; note that Schr\"odinger did not discuss this issue in his original paper. Here the analysis is also similar to the bound-state problem. If the branch cuts run to infinity, none of the contours that run to infinity yield finite wavefunctions because they all diverge as $r\to 0$. We find the only acceptable contour arises when we draw the branch cut from $c_1$ to $c_2$ (here along the imaginary axis) and we pick the contour $\gamma_5$ to run in the counter-clockwise direction around both $c_1$ and $c_2$ as shown in Fig.~\ref{fig:cont_5}. Note that the exponents are never integers here, so the only choice that will work must be  a closed contour; Hankel contours suffer from the same issues we saw with the bound states and have the wavefunction diverge as $r\to 0$.

One can immediately verify that when we choose the contour around the branch cut, $\lim_{r\to 0} U_5(r)$ satisfies
\begin{equation}
    \lim_{r\to 0}U_5(r)=\oint_{\gamma_5} dz\,(z-c_1)^{\alpha_1-1}(z-c_2)^{\alpha_2-1}.
\end{equation}
By deforming the contour tight around the branch cut (dashed contour), one immediately sees that the two circular integrals around $c_1$ and $c_2$ vanish, because they depend on the infinitesimal radius with a positive integer power $\epsilon^{l+1}$. The remainder of the integral is given by a path along the vertical lines that run up from $c_1$ to $c_2$ on one side of the branch cut and then back down on the other. The phases for the branch cut are $\pi/2$ for $(z+ic)^{\alpha_1-1}$ and $-\pi/2$ for $(z-ic)^{\alpha_2-1}$ on the right side of the branch cut and $-3\pi/2$ for $(z+ic)^{\alpha_1-1}$ and $3\pi/2$ for $(z-ic)^{\alpha_2-1}$ on the left side of the branch cut.
This gives
\begin{eqnarray}
    \lim_{r\to 0}U_5(r)&=e^{i\frac{\pi}{2}(\alpha_1-\alpha_2)}(2c)^{2l+1}\int_0^1 dx\,x^{\alpha_1-1}(1-x)^{\alpha_2-1}\nonumber\\
    &+e^{-i\frac{3\pi}{2}(\alpha_1-\alpha_2)}(2c)^{2l+1}\int_0^1 dx\,x^{\alpha_2-1}(1-x)^{\alpha_1-1}\nonumber\\
    &=e^{i\frac{\pi}{2}(\alpha_1-\alpha_2)}(2c)^{2l+1}B(\alpha_1,\alpha_2)\left (1+e^{-2i\pi(\alpha_1-\alpha_2)}\right ).
\end{eqnarray}
Here, $B(x,y)$ is the beta function. Using the values for $\alpha_1$ and $\alpha_2$, we find $B(\alpha_1,\alpha_2)=|\Gamma(l+1+i\hbar/(a_0\sqrt{2\mu E}))|^2/\Gamma(2l+2)$, which is finite for all $l$.  The key for this calculation is that the wavefunction is well-defined and finite everywhere and thereby yields the continuum solution. A more careful analysis shows that the continuum wavefunction is also real. The treatise by Bethe and Salpeter~\cite{bethe_salpeter} discusses this solution to a limited extent. They do not provide a detailed analysis, but do present the final results.

Schr\"odinger did discuss the asymptotic behavior as $r\to\infty$, finding these functions behave like $\exp(\pm icr)r^{-l-1}$, which implies that the functions $\chi(r)=r^lU_5(r)$ all behave asymptotically as $\exp(\pm icr)/r$. This is easy to show by two changes of variables. The wavefunction is given by
\begin{equation}
    U_5(r)=\oint_{\gamma_5}dz\, e^{rz}(z+ic)^{\alpha_1-1}(z-ic)^{\alpha_2-1}.
\end{equation}
The asymptotic analysis requires just two steps. We illustrate it for the contributions that come from the integral around $c_2$: first, shift $z\to z+ic$, which gives us
\begin{equation}
U_5(r)=e^{icr}\oint dz\,e^{rz}(z+2ic)^{\alpha_1-1}z^{\alpha_2-1};
\end{equation}
second, rescale $z=w/r$ to yield
\begin{equation}
    U_5(r)=e^{icr}\frac{1}{r^{\alpha_2}}\oint dw\,e^{w}\left (\frac{w}{r}+2ic\right )^{\alpha_1-1}w^{\alpha_2-1}.
\end{equation}
The asymptotic behavior with respect to $r$ has now emerged from the integral because ${\rm Re}(\alpha_2)=l+1$. The remaining integral is approximately constant, and is proportional to $\Gamma(\alpha_2)$ in the limit as $r\to\infty$. A similar analysis can be done for the contributions coming from the region around $c_1$. One can easily see now how to verify Schr\"odinger's claims about the asymptotic behavior. 

We have now determined the continuum wavefunction, but it is expressed as a contour integral; this is because contour integrals around two branch points connected by a branch cut usually do not have simple alternative analytic expressions. This contour integral can be related to the confluent hypergeometric function of a complex argument or one can determine power series expansions that approximate the wavefunction using standard methods. We do not go into further details of this, as they were not discussed by Schr\"odinger and the results are well known~\cite{bethe_salpeter}.

One final point remains, we could, in principle, have defined the solution via an integral that runs from $c_1$ to $c_2$ only (rather than surrounding it as $\gamma_5$ does) for the cases where $l\ne 0$. Because such a function has the same asymptotic behavior as $r\to 0$ and $r\to\infty$, it must be proportional to the solution we did employ with the contour enclosing both branch points, so we do not discuss this issue further here.

\section{Conclusions}

In this work, we describe how one can employ the Laplace method for solving differential equations to determine the quantum-mechanical wavefunctions of hydrogen. The methodology requires an intermediate knowledge of complex analysis that includes how to define branch cuts, what the definitions of the logarithm and $\Gamma$ function are, and how to determine the asymptotic behavior of contour integrals via the techniques of steepest descents and stationary phase. Much of this material can be taught within the quantum classroom if one wants to move away from the standard Frobenius method for solving differential equations. We feel that this is  worthwhile,  even if it may take substantial time, because there are many fields of physics that require advanced knowledge of complex analysis and this is a good opportunity to incorporate it within the quantum-mechanics curriculum. Most likely this would be done at the graduate level. In addition, we showed in detail just how Schr\"odinger solved the original hydrogen problem, including all of the technical details omitted in the original work. It is important to make sure that this critical scientific achievement does not become a lost art. We conclude with a comment that the application of the Laplace method need not end with just the solution of hydrogen. It can be also applied to essentially all of the analytically solvable problems in quantum mechanics.

\section*{Acknowledgments}
This work was supported by the National Science Foundation under grant number PHY-1915130. In addition, JKF was supported by the McDevitt bequest at Georgetown University. AG acknowledges support through Schr\"odinger fellowship J-4267 of the Austrian Science Fund (FWF) and through a 'Sub auspiciis Exzellenzstipendium' of the  Austrian Federal Ministry of Education, Science and Research. We also want to thank Konstantin Tikhonov and Yaroslav Rodionov, whose edX course MISiSx:18.11x ``Complex Analysis with Physical Applications'' inspired us to initiate this project.

\end{document}